\documentclass[a4paper]{article}
\usepackage[top=2.8cm,bottom=3cm,left=3cm,right=3cm]{geometry}
\usepackage{graphicx}
\usepackage{txfonts}
\usepackage{amsfonts}
\usepackage{amssymb}
\usepackage{booktabs}
\usepackage{xcolor}
\usepackage{lineno}
\usepackage{natbib}
\bibpunct{(}{)}{;}{a}{}{,} 
\begin{document}
\Large
\textbf{\begin{center}{RIS: Regularized Imaging Spectroscopy \\ for STIX on-board Solar Orbiter} \end{center}}

\normalsize

Anna Volpara$^1$, Alessandro Lupoli$^2$, Frank Filbir$^{2,3}$, Emma Perracchione$^4$, Anna Maria Massone$^{1}$, and Michele Piana$^{1,5}$ \\

\hspace{-0.5cm}$^1$ MIDA, Dipartimento di Matematica, Universit\`a di Genova, via Dodecaneso 35 16146 Genova, Italy  \\
email: volpara@dima.unige.it, massone@dima.unige.it, piana@dima.unige.it \\
$^2$ Department of Mathematics, TUM School of Computation, Information and Technology, Technische Universität München, Boltzmannstraße 3 85748 Garching b. München  \\
email: frank$\_$filbir@web.de, lupa@ma.tum.de \\
$^3$ Bioengineering Center, Institute of Biological and Medical Imaging, Helmholtz Munich, Ingolstädter Landstraße 1
85764 Neuherberg, Germany \\
$^4$ Dipartimento di Scienze Matematiche "Giuseppe Luigi Lagrange", Politecnico di Torino, Corso Duca degli Abruzzi 24 10129 Torino, Italy \\
email: emma.perracchione@polito.it \\
$^5$ Istituto Nazionale di Astrofisica, Osservatorio Astrofisico di Torino, via Osservatorio 20 10025 Pino Torinese Italy \\

%
\date{\today}

\begin{center}
   \textbf{Abstract} 
\end{center}
   Imaging spectroscopy, i.e., the generation of spatially resolved count spectra and of cubes of count maps at different energies, is one of the main goals of solar hard X-ray missions based on Fourier imaging. For these telescopes, so far imaging spectroscopy has been realized via the generation of either count maps independently reconstructed at the different energy channels, or electron flux maps reconstructed via deconvolution of approximate forms of the bremsstrahlung cross-section.
   
   Our aim is to introduce the Regularized Imaging Spectroscopy method (RIS), in which regularization implemented in the count space imposes a smoothing constraint across contiguous energy channels, without the need to compute any de-convolution of the bremsstrahlung effect.
  
   STIX records imaging data computing visibilities in the spatial frequency domain. Our RIS is a sequential scheme in which part of the information coded in the image reconstructed at a specific energy channel is transferred to the reconstruction process at a contiguous channel via visibility interpolation computed  by means of Variably Scaled Kernels.
  
   In the case of STIX visibilities recorded during the November 11, 2022 flaring event, we show that RIS is able to generate hard X-ray maps whose morphology smoothly evolves from one energy channel to the contiguous one, and that, accordingly, from these maps it is possible to infer spatially-resolved count spectra characterized by notable numerical stability. We also show that the performances of this approach are rather robust with respect to both the image reconstruction method and the count energy channel utilized to trigger the sequential process.
 
   We conclude that RIS is appropriate to construct image cubes from STIX visibilities that are characterized by a smooth behavior across count energies, thus allowing the generation of numerically stable (and, thus, physically reliable) local count spectra.

\textbf{key words.} Techniques: imaging spectroscopy; Sun: X-rays, gamma rays; Sun: flares; Methods: numerical



\section{Introduction}\label{introduction}
Imaging spectroscopy in solar hard X-ray imaging \citep{piana2022hard} consists in providing, on the one hand, reconstructions of hard X-ray count maps at different count energy channels and, on the other hand, spatially-resolved count spectra. From a scientific viewpoint, imaging spectroscopy allows inferring information about the characteristics of the location of the acceleration region \citep{2006A&A...456..751B,2008ApJ...673..576X,2011SoPh..269..269M,2012A&A...543A..53G} and even studying the physics of electron transport within the source \citep{2010ApJ...712L.131P,2024A&A...684A.185V}.

In the last decades, imaging spectroscopy has been one of the main science goals of space missions that have been designed to generate measurements representing samples of the Fourier transform of the incoming radiation flux, named visibilities, for energy values in the range between a few keV and some tenths of keV \citep{enlighten1658,2002SoPh,krucker2020spectrometer,2019RAA....19..160Z,2023SoPh..298..114M}. In the case of these Fourier-based imagers, the traditional approach to imaging spectroscopy applies Fourier inversion methods to sets of visibilities, to obtain spatial images at different count energies \citep{2006A&A...456..751B,2008ApJ...673..576X,2011SoPh..269..269M}. This approach is simple, but not optimal. Indeed, each image is reconstructed using a visibility set that is completely independent of those used to make images in adjacent energy bins, which implies that reconstructions corresponding to adjacent count energy bins are uncorrelated and may present substantial differences from each other. Therefore, stacking these images leads to nonphysical fluctuations in the spatially resolved count spectra, which makes their interpretation problematic.

To address this issue, a more sophisticated method has been developed within the RHESSI and STIX frameworks, in which the observed visibility spectra obtained at points in the spatial frequency domain were inverted by means of a regularization method providing electron flux visibility spectra smoothed along the electron energy direction \citep{2007ApJ...665..846P,prato2009regularized,2024A&A...684A.185V}. These latter ones can be utilized to both reconstruct electron flux images and, once projected back to the count space, regularized count spectra. By construction, these spectra vary smoothly along the count energy direction. However, the main weakness of this approach is in its computational complexity. In fact, in this case spatially resolved regularized count spectra are obtained by first inverting the original experimental visibility spectra, which is an intrinsically ill-posed problem \citep{bertero2021introduction}; further, to do so, a specific model for the bremsstrahlung cross-section must be assumed, which requires an approximation of the electron-ion interaction process \citep{1959RvMP...31..920K}; finally, the regularized count spectra are obtained by means of a back-projection onto the count space, which requires the use of numerical integration \citep{2024A&A...684A.185V}.

The objective of the present paper is to introduce an approach to imaging spectroscopy for visibility-based hard X-ray telescopes that addresses the two previous limitations; i.e., this approach aims to introduce a regularization constraint across the count energies without any spectral inversion process, and therefore without the need to previously construct any electron flux maps. The starting point of the {\em{Regularized Imaging Spectroscopy method (RIS)}} is an interpolation/extrapolation method based on the use of Variably Scaled Kernels (VSKs), which allow an advanced approach to numerical approximation where a priori information can be easily plugged into the interpolation process by means of appropriate scale functions \citep{bozzini2015interpolation}. In the case of both RHESSI and STIX, VSKs proved to be very effective in interpolating visibilities in the frequency plane, while a non-linear iterative scheme \citep{piana1997projected} applied to the interpolated visibility surface was able to provide physically reliable reconstructions of the flaring source \citep{2021ApJ...919..133P}. 

In the present context we decided to use VSKs and iterative image reconstruction as the computational engine of a regularized imaging spectroscopy approach that can be formulated according to the following scheme:
\begin{enumerate}
    \item Given a visibility set at a starting {\em{triggering count energy channel}} $\epsilon_0$, apply any visibility-based image reconstruction method to obtain the corresponding {\em{triggering map}}.
    \item Apply Fourier transform to compute the scale function corresponding to the reconstruction obtained in the previous step.
    \item Given the visibility set in an energy channel $\epsilon_1$ adjacent to $\epsilon_0$, apply VSK with the scale function obtained in the previous step to realize interpolation in the spatial frequency plane.    
    \item Apply the constrained iterative method in order to obtain the reconstruction of the count map at energy $\epsilon_1$.
    \item Repeat the process up to a given final energy channel $\epsilon_I$.
\end{enumerate}
Of course, the actual implementation of this scheme, and, consequently, its effectiveness in reproducing a smooth spectral evolution of the flaring morphology, may depend on the choice of the count energy channel $\epsilon_0$ and of the visibility-based image reconstruction method used at step 1 to trigger the process. After presenting the mathematical description of RIS, the present study validates its performances in the case of experimental STIX visibilities recorded in correspondence with the November 11, 2022 event, also assessing the impact on the reconstructions' reliability of different choices of the triggering energy and image reconstruction method.

The plan of the paper is as follows. Section 2 illustrates the mathematical setup of VSK-based regularized imaging spectroscopy, and Section 3 describes applications against STIX observations. Our conclusions are offered in Section 4.

\section{Mathematical formulation of the method}\label{mathematical-setup}
The STIX image reconstruction problem can be written as
\begin{equation}\label{eq:signal-formation-processing}
 {\bf{V}}={\bf{A}}{\bf{f}} \ ,   
\end{equation}
where ${\bf{A}}$ is the the discretized version of the Fourier transform sampled at a set of points $\{{\bf{u}}_k = (u_k,v_k)\}_{k=1}^{n}$ inside a disk $B$ in the $(u,v)$ plane; ${\bf{V}}$ is the vector whose $n$ components are the complex experimental visibilities observed by the telescope at a specific energy channel $\epsilon_0$ and in a specific time window. Finally, ${\bf{f}}$ is the vector whose components are the discretized values
of the incoming radiation flux. Any visibility-based imaging method provides an (either linear or non-linear) estimate ${\cal{F}}_{\alpha}$ of ${\bf{A}}^{-1}$ such that
\begin{equation}\label{eq:reconstructed-image}
    {\bf{f}}_{\alpha} = {\cal{F}}_{\alpha}({\bf{V}})
\end{equation}
is the reconstructed image at energy channel $\epsilon_0$. The subscript $\alpha$ denotes the fact that image reconstruction relies on some regularization process, tuned by means of a regularization parameter $\alpha$. Regularization reduces the numerical instabilities introduced by the limited number of Fourier samples at disposal. However, here this regularization has just a spatial meaning, and does not introduce any smoothing effect along the spectral direction. As a consequence of this, the reconstruction provided by ${\cal{F}}_{\alpha}$ when applied to the visibility set recorded by STIX at a contiguous energy interval $\epsilon_1$ is totally uncorrelated with respect to the reconstruction provided by the same method at $\epsilon_0$, which is significantly unreliable from a physical viewpoint. 

The present study shows that a smoothing constraint in the spectral direction for imaging spectroscopy can be introduced by using numerical interpolation in the spatial frequency plane based on radial basis functions (RBFs) and variably scaled kernels (VSKs). According to what described in \cite{2021ApJ...919..133P,2021InvPr..37j5001P}, this approach constructs the interpolant of the visibilities
\begin{equation}\label{eq:interpolation-1}
    P({\bf{u}}_k) = {\bf{V}}_k\ \ \ k=1,\ldots,n \ ,
\end{equation}
where
\begin{equation}\label{eq:interpolation-2}
    P({\bf{u}}) = \sum_{k=1}^{n}a_k b_k({\bf{u}}) \ ,
\end{equation}
and
\begin{equation}\label{eq:RBF-scale}
    b_k({\bf{u}}) = \phi(\|({\bf{u}},\psi({\bf{u}})) - ({\bf{u_k}},\psi({\bf{u}_k}))\|) \ , \ k=1,\ldots,n.
\end{equation}
In equation (\ref{eq:RBF-scale}), $\phi$ is an appropriately chosen RBF, $\| \cdot \|$ is the Euclidean norm, and $\psi$ is a scale function encoding prior information on the flaring source. Once $\phi$ and $\psi$ have been chosen, the numerical solution of equations (\ref{eq:interpolation-1})-(\ref{eq:RBF-scale}) leads to the determination of the coefficients $\{a_k\}_{k=1}^n$ and then to the computation of the interpolated visibility surface ${\tilde{\bf{V}}}$ such that
\begin{equation}\label{visibility-surface}
   {\tilde{\bf{V}}}_j =  P({\tilde{{\bf{u}}}}_j) \ ,  \ j=1,\ldots,N^2 \ ,
\end{equation}
where the knots $\{{\tilde{{\bf{u}}}}_j\}_{j=1}^{N^2}$ are picked up on a regular mesh in the $(u,v)$ plane with $N^2 \gg n$. Therefore, interpolation leads from equation (\ref{eq:signal-formation-processing}) to 
\begin{equation}\label{eq:signal-formation-processing-new}
    {\tilde{\bf{V}}} = {\tilde{\bf{A}}}{\tilde{{\bf{f}}}} \ ,
\end{equation}
where ${\tilde{\bf{A}}}$ is the Fourier transform now discretized over an $N^2 \times N^2$ grid, ${\tilde{\bf{V}}}$ is the $N^2 \times 1$ vector containing all the interpolated visibilities, and ${\tilde{{\bf{f}}}}$ is the $N^2 \times 1$ solution vector. The inversion of equation (\ref{eq:signal-formation-processing-new}) can be performed by means of several regularization methods, although the projected Landweber iterative scheme \citep{piana1997projected,2021InvPr..37j5001P} proved to have super-resolving powers, thanks to the iterative application of the convex projector onto the set of vectors with non-negative components.

\subsection{RIS: Regularized Imaging Spectroscopy in the count space}
VSK-based interpolation/extrapolation in the visibility domain is the computational tool that allows the realization of RIS in the count space, i.e., without the need to construct the electron maps via bremsstrahlung de-convolution. 

Given an input visibility vector ${\bf{V}}_{\epsilon_0}$ at a specific triggering count energy channel $\epsilon_0$, the initialization step first applies any Fourier-based image reconstruction to generate the triggering reconstructed image ${\bf{f}}_{\epsilon_0}$. The Fourier transform of ${\bf{f}}_{\epsilon_0}$ is chosen as scaling function $\psi_{\epsilon_0}$. Given the visibility set at $\epsilon_1$ (which is contiguous to $\epsilon_0$), VSK-based interpolation with $\psi_{\epsilon_0}$ as scale function computes the visibility surface $ {\tilde{\bf{V}}}_{\epsilon_1}$ and the application of the projected iterative scheme to obtain ${\bf{f}}_{\epsilon_1}$ concludes the initialization step. 

\begin{figure}
     \centering
     \includegraphics[width=0.9\textwidth]{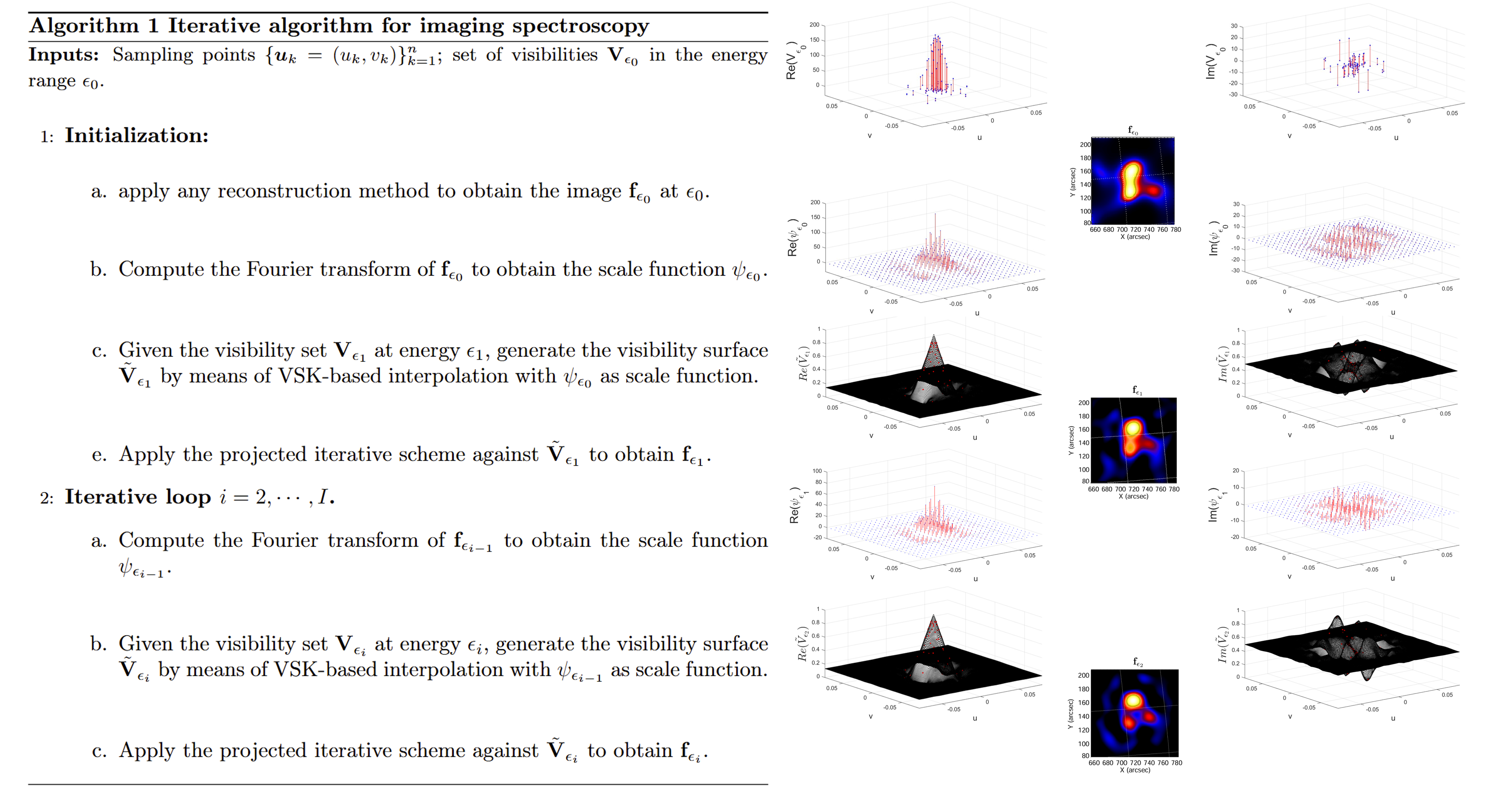}
 \caption{The regularized imaging spectroscopy algorithm and the outcomes corresponding to each one of its steps.}
     \label{fig:fig-1}
 \end{figure}

Now, the sequential loop is made over a set of contiguous count energies $\{\epsilon_i \}_{i=2}^I$, in such a way that the Fourier transform of ${\tilde{\bf{f}}}_{\epsilon_{i-1}}$ at energy $\epsilon_{i-1}$ is used as scale function for the generation of the visibility surface at energy $\epsilon_i$, and the new ${\tilde{\bf{f}}}_{\epsilon_{i}}$ is computed by means of the usual projected iterative scheme. Therefore, this approach realizes regularization along the spectral direction by exploiting the scale function to transfer some topographical information from two contiguous energy channels. A pictorial description of this scheme is illustrated in Figure \ref{fig:fig-1}, which also contains the outcome of each step of the algorithm.

From a technical viewpoint, the application of RIS to STIX observations shown in the next section has been realized by means of an implementation that utilizes the Matern $C^0$ kernel
\begin{equation}\label{eq:matern}
\phi(\|u\|) = e^{-\eta \|u \|}
\end{equation}
as RBF, where $\eta$ is the so-called shape parameter. Further, the stopping rule for the projected Landweber scheme is based on a check of the $\chi^2$ values, and the initialization vector is ${\bf{f}}^{(0)}_{\epsilon_i}=0$ for each $i=1,\ldots,I$. Further, the sequential image reconstruction is interrupted at the energy channel where the signal-to-noise ratio is too low to guarantee a reliable count map. Finally, as far as the measurements are concerned, in the applications below we have utilized eight of the ten STIX detectors (this because the two detectors with highest resolution are not yet completely calibrated). This corresponds to a visibility set made of $48$ experimental visibilities overall, where $24$ visibilities are measured by STIX and the other $24$ are generated accounting for the fact that $V(-u,-v)$ is equal to the complex conjugate of $V(u,v)$. Further, the disk $B$ has a radius of $0.03$ arcsec$^{-1}$.

\section{Applications to STIX visibilities}\label{applications}
\begin{figure}
 \centering \includegraphics[width=0.75\textwidth]{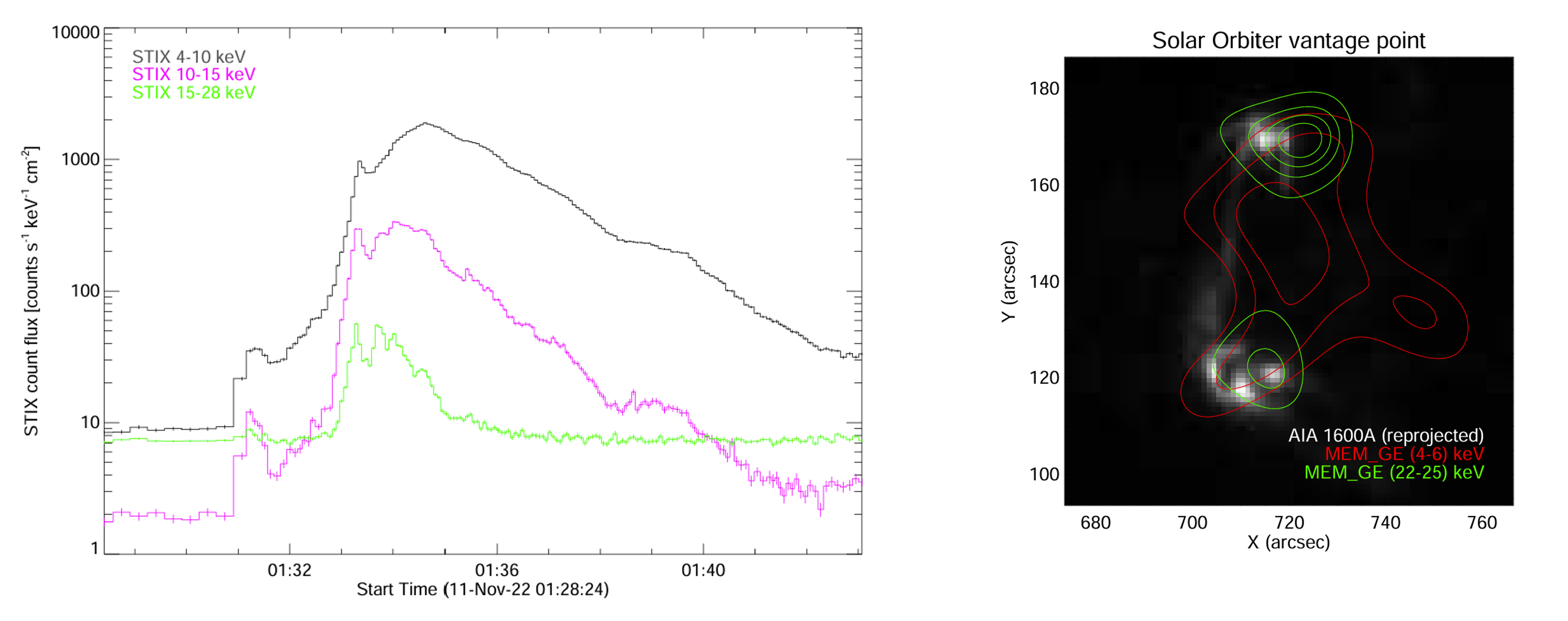} 
 \caption{Hard X-ray emission of the November 11, 2022 event in the time window between 01:30:00 and 01:32:00 UT. Left panel: STIX light-curves corresponding to three energy channels. Right panel: Level curves of the count emission provided by MEM$\_$GE and superimposed to the $1600 \AA$ emission recorded by SDO/AIA  (the AIA data have been appropriately reprojected in order to account for the Solar Orbiter and SDO different vantage points). The red level curves correspond to the thermal emission in the energy range $4-6$ keV, while the green ones correspond to the non-thermal emission in the energy range $22-25$ keV.}\label{fig:lightcurves}
 \end{figure}

In order to show the performances of RIS in the case of STIX data, we considered the November 11, 2022 event in the time window between 01:30:00 and 01:32:00 UT. The sequence of energy channels was made of $9$ $2$-keV-wide channels starting from $4-6$ keV to $20-22$ keV, plus one $3$-keV-wide channel at $22-25$ keV. Figure \ref{fig:lightcurves} shows some light-curves corresponding to the event, and some level curves of the STIX map at $4-6$ keV (in red) and $22-25$ keV (in green) as superimposed on the EUV map recorded by the Atmospheric Imaging Assembly on-board the Solar Dynamics Observatory (SDO/AIA). The reconstruction method utilized for generating these level curves is a constrained maximum entropy method implemented in the MEM$\_$GE routine of the STIX ground software \citep{2020mem_ge}.

\begin{figure}
     \centering
     \includegraphics[width=0.75\textwidth]{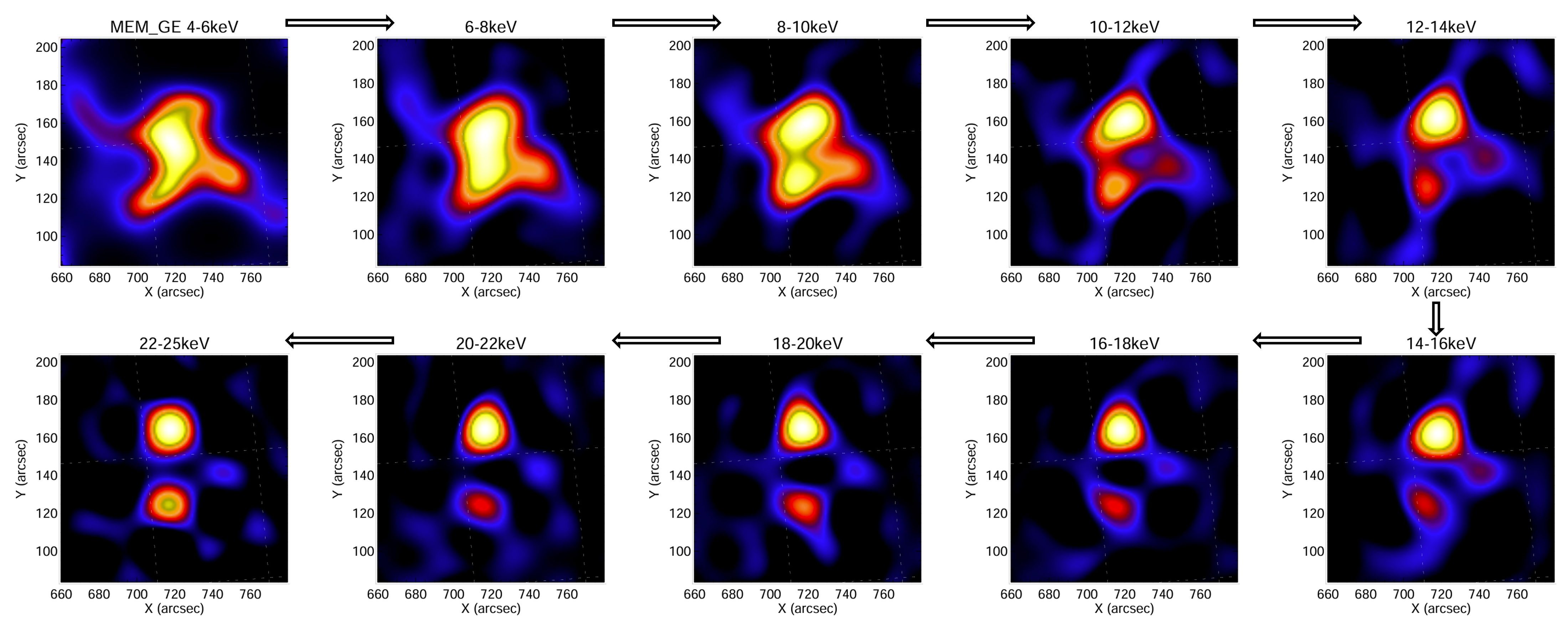}\\
     \vspace{1cm}
     \includegraphics[width=0.75\textwidth]{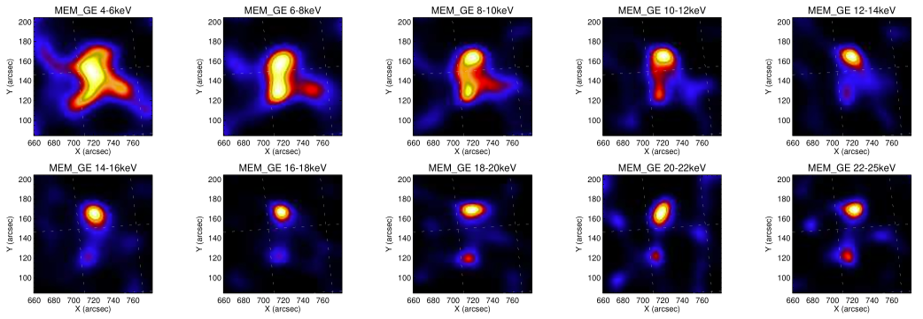}
 \caption{Comparison between the maps provided at contiguous energy channels by the regularized imaging spectroscopy method and the reconstructions provided by MEM$\_$GE at the same channels. Top part: reconstructions provided by RIS using the MEM$\_$GE reconstruction at the top left panel as triggering image. Bottom part: reconstructions provided by MEM$\_$GE.}\label{fig:regularized-IS}
 \end{figure}

The two top rows of Figure \ref{fig:regularized-IS} contain the count maps corresponding to the November 11, 2022 event provided by RIS. The reconstruction used to trigger the scheme is the one provided by MEM$\_$GE in correspondence of the visibility set recorded by STIX at $4-6$ keV (see the map in the top left panel in the first row of the figure). Here the regularized reconstructions are compared to the reconstructions provided by MEM$\_$GE in the two bottom rows, from the same visibility sets but without any correlation constraint across consecutive energy channels. The comparison clearly points out that regularization nicely implies a smooth and realistic evolution of the flaring morphology, while the uncorrelated maps suffer some numerical instability particularly at high energies, where the signal-to-noise ratio of STIX counts significantly decreases.

\begin{figure}
     \centering
     \includegraphics[width=0.65\textwidth]{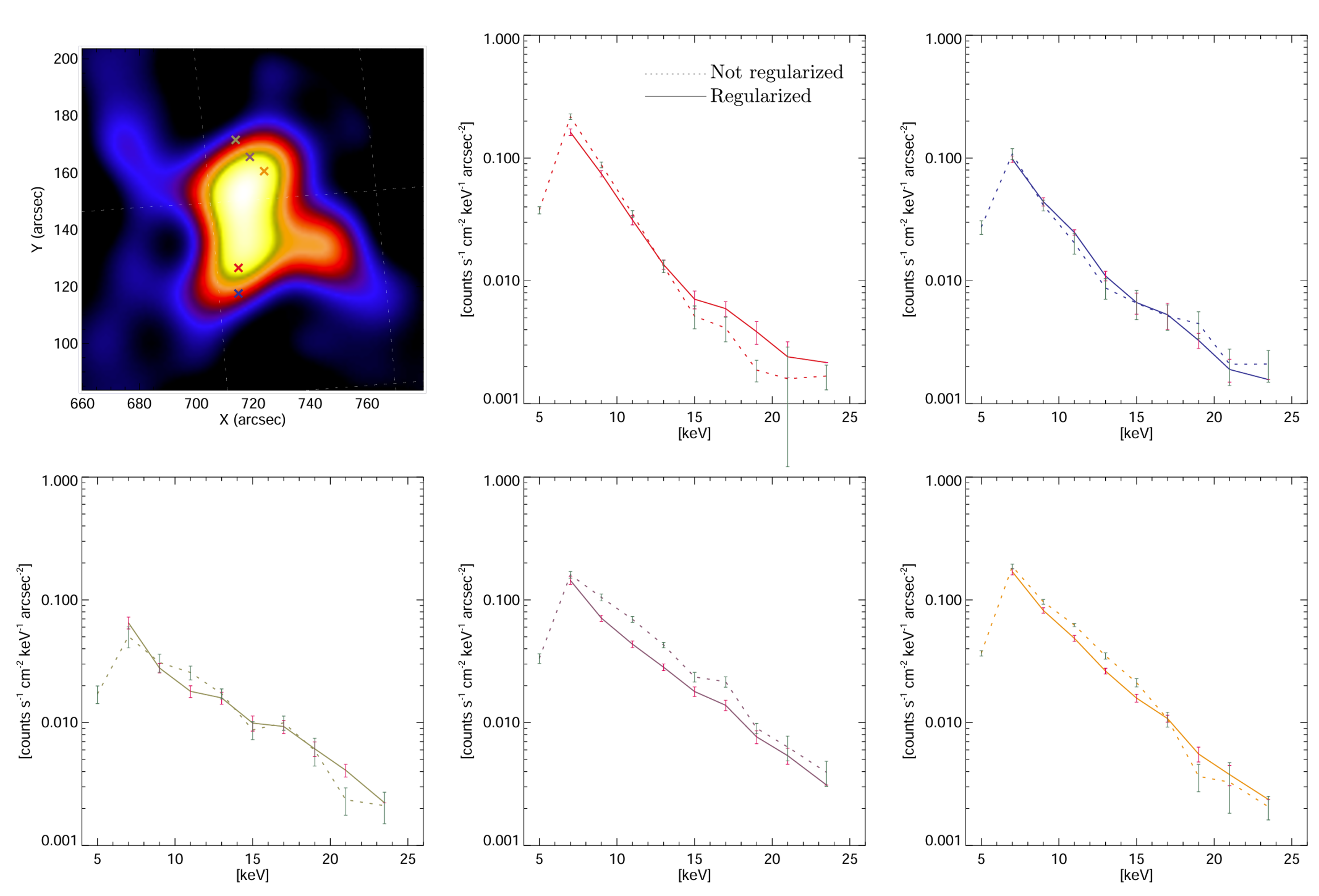}
 \caption{Local count spectra corresponding to the points highlighted by the colored crosses superimposed to the map at the top left panel. The local spectra obtained by using the maps in Figure \ref{fig:regularized-IS} provided by RIS are in solid line, while the ones obtained by using the maps in the same figure but provided by MEM$\_$GE when independently applied at each energy channel are in dashed line. }\label{fig:local-spectra}
 \end{figure}

The main advantage of this approach to imaging spectroscopy is in the possibility to reconstruct numerically stable spatially-resolved count spectra. This potentiality is illustrated in Figure \ref{fig:local-spectra}, in which local count spectra obtained from the regularized maps in Figure \ref{fig:regularized-IS} are compared to the ones obtained from maps reconstructed by MEM$\_$GE, independently at each one of the same energy channels. The points in the flaring region at which these local spectra are computed, are identified by the colored crosses in the map at the top left panel of the figure. Once again, regularization implies numerical stability and smaller uncertainties, as demonstrated by error bars that are systematically smaller than the ones characterizing the not-regularized spectra.

\begin{figure}
     \centering
     \includegraphics[width=0.8\textwidth]{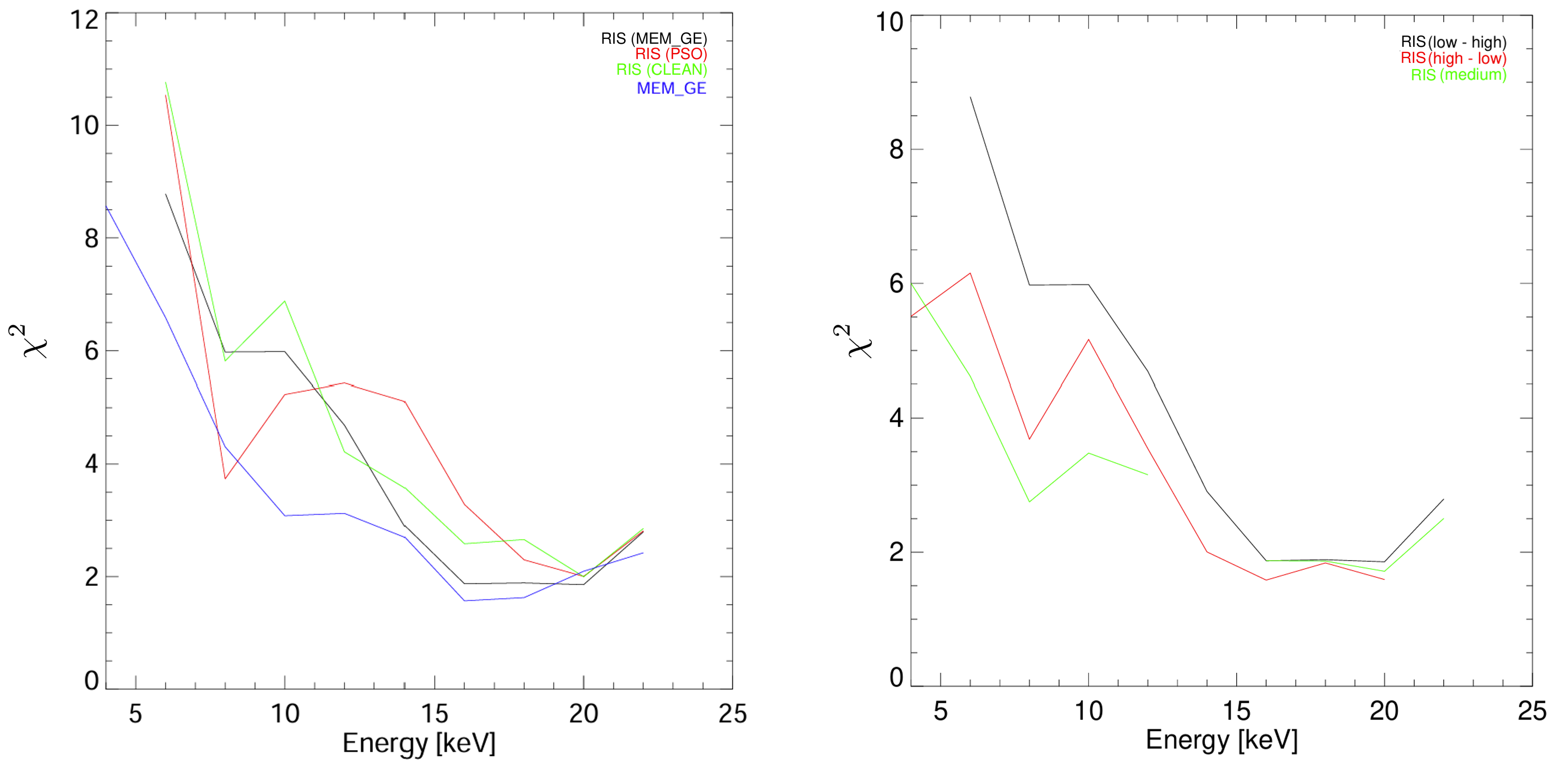}
 \caption{Robustness of RIS with respect to the reconstruction method applied to generate the triggering image and on the triggering energy. Left panel: $\chi^2$ values at different energies when the triggering image is reconstrcuted by means of MEM$\_$GE (black), forward-fit with PSO (red) and CLEAN (green). The $\chi^2$ values provided by MEM$\_$GE independently at each energy channel are shown in blue. Right panel: $\chi^2$ values at different energies when the triggering energy corresponds to the lowest channel and the regularization process flows from low to high energies (black), the triggering energy corresponds to the highest channel and the regularization process flows from high to low energies (red), and the triggering energy has an intermediate value with the regularization process flowing along both directions (green).}\label{fig:robustness}
 \end{figure}

RIS can be triggered by images reconstructed by means of different image reconstruction methods, and can start its sequential process from any energy channel (the only condition is that the energy channels involved in the process are contiguous). Therefore, one of the possibile validation issues related to this approach is concerned with robustness with respect to these two degrees of freedom. The left panel of Figure \ref{fig:robustness} compares the $\chi^2$ values computed from RIS maps generated by using triggering images reconstructed by MEM$\_$GE, CLEAN \citep{Clean}, and a forward-fit algorithm realized by means of Particle Swarm Optimization (PSO) \citep{2022A&A...668A.145V}. In all these cases the triggering energy is the $4-6$ keV energy channel.  In order to show the reliability of RIS maps, the left panel in the figure also shows the $\chi^2$ values obtained from maps provided by the MEM$\_$GE algorithm with no correlation constraint across the energies. The right panel does the same in the case of three different choices of the triggering energy channel when, in all cases, the triggering image is the ones provided by MEM$\_$GE. Specifically, the triggering energies are $\epsilon_0=4-6$ keV (i.e., regularization proceeds from low to high energies);  $\epsilon_0=14-16$ keV, i.e. regularization starts at an intermediate energy and from that proceeds to both low and high energies; $\epsilon_0 = 22-25$ keV, i.e. regularization proceeds from high to low energies. In all these cases, the $\chi^2$ values are rather stable and decreases consistently from low to high energies, given that $\chi^2$ is normalized with respect to the experimental standard deviation, which is smaller at low energies. Interestingly, the best fit associated to the regularization flow from high to low energies is more accurate than the one associated to the low-to-high-energy flow, which has a rather immediate physical interpretation. Indeed, bremsstrahlung implies that high energy photons (and the corresponding counts) are produced by a small number of electrons. As a consequence, when the triggering energy is high, regularization flows along the same direction of the increase of the physics-induced uncertainty.

\section{Conclusions}\label{conclusions}
RIS is a regularization method for the reconstruction of count maps from STIX visibilities at different energy channels, which are characterized by a smooth behavior across such energies. This smoothing constraint is imposed by means of a sequential scheme based on the use of VSK interpolation in the spatial frequency domain. The main advantage of this approach is that, from these regularized maps, it is possibile to produce spatially-resolved count spectra characterized by unprecedented numerical stability. RIS is also robust with respect to the choice of both the reconstruction method applied to generate the image that triggers the regularization process, and the triggering energy channel. 

In our opinion, next research concerning RIS may have two possible developments. First, from a physical viewpoint, the systematic use of regularized local count spectra will allow a spectroscopy-based and reliable description of the interplay of thermal and non-thermal mechanisms in solar flares at different locations in the flaring region. Second, from a methodological viewpoint, the RIS approach might be extended to a regularized approach to dynamical studies of the flaring emission, where the VSK-based smoothing constraint will be realized across contiguous time intervals.

\section*{Acknowledgements}
Solar Orbiter is a space mission of international collaboration between ESA and NASA, operated by ESA. The STIX instrument is an international collaboration between Switzerland, Poland, France, Czech Republic, Germany, Austria, Ireland, and Italy. 
AV, MP, and AMM acknowledge the support of the “Accordo ASI/INAF Solar Orbiter: Supporto scientifico per la realizzazione degli strumenti Metis, SWA/DPU e STIX nelle Fasi D-E”, and the HORIZON Europe ARCAFF Project, grant No. 101082164. MP also acknowledges financial support under the National Recovery and Resilience Plan (NRRP), Mission 4, Component 2, Investment 1.1, Call for tender No. 104 published on 2.2.2022 by the Italian Ministry of University and Research (MUR), funded by the European Union – NextGenerationEU– Project Title 'Inverse Problems in Imaging Sciences (IPIS)" – CUP D53D23005740006 - Grant Assignment Decree No. 973 adopted on 30/06/2023 by the Italian Ministry of Ministry of University and Research (MUR).

\bibliographystyle{aa}
\bibliography{bib_stix_arxiv}

\end{document}